# Magnetothermodynamics of the Ising Antiferromagnet $Dy_2Ge_2O_7$

X. Ke[1], M. L. Dahlberg[1], E. Morosan[2], J.A. Fleitman[3], R. J. Cava[3], and P. Schiffer[1]

1. Department of Physics and Materials Research Institute, Pennsylvania State University, University Park, PA 16802 USA.

2. Department of Physics and Astronomy, Rice University, Houston, TX 77005 USA.

3. Department of Chemistry and Princeton Materials Institute, Princeton University, NJ, 08540 USA.

We report systematic low temperature measurements of the DC magnetization, AC susceptibility, and heat capacity of dysprosium pyrogermanate ($Dy_2Ge_2O_7$) single crystal and powder samples. Our results confirm that $Dy_2Ge_2O_7$ is an anisotropic antiferromagnet. The isothermal field dependent magnetization and the integrated magnetic entropy both indicate that the $Dy^{3+}$ ions behave as Ising-like spins, analogous to those in the pyrochlore spin ice materials. Both single-spin and collective spin relaxation phenomena appear to lead to spin freezing in this material, again in analogy to observations in the spin ice materials, suggesting that such phenomena may be generic to a broader class of magnetic materials.

PACS numbers: 75.30.Cr, 75.30.Gw, 75.40.Cx, 75.40.Gb.

In the past decade, there has been increasing interest in the rare earth pyrochlore oxide materials ($A_2B_2O_7$) as models in which to study geometric frustration of the interactions between the magnetic moments of rare earth ions in a regular lattice [1,2]. Rare earth pyrochlores with titanium and tin on the B site are of particular interest; depending on the rare earth ions residing on the A sites, these materials display a variety of ground states [3] including unconventional long range order, 'spin liquid' states, and the 'spin ice' states [4,5,6,7,8,9,10].

Rare-earth based pyrogermanates ($Ln_2Ge_2O_7$), with a similar chemical formula, but a different structure from the pyrochlores, have also attracted considerable attention [11,12,13,14,15,16,17]. Instead of having a cubic pyrochlore structure with a sublattice of magnetic corner-sharing tetrahedra, rare-earth pyrogermanates are reported to have either triclinic (for Ln = La, Pr, or Nd-Gd), tetragonal (for Ln = Gd-Lu), or monoclinic structures ($Sc_2Ge_2O_7$) [11]. DC magnetic susceptibility measurements of Tb, Dy, and Ho-based pyrogermanate single crystals show magnetic anisotropy resulting from strong crystal field effects [12,13,14]. In this paper, we explore the low temperature properties of $Dy_2Ge_2O_7$, including DC magnetization, AC susceptibility, and heat capacity measurements. This compound develops long range antiferromagnetic order with a Néel temperature ($T_N$) on the order of 2 K, which is manifested by strong features in all measured properties. The AC susceptibility well above $T_N$ indicates the presence of a single-ion spin relaxation similar to that observed in $Dy_2Ti_2O_7$ [18,19], $Dy_2Sn_2O_7$ [8] and $Ho_2Ti_2O_7$ [20] spin ice systems, and correlated spin relaxation at low temperatures. In contrast to the spin ice materials, in which a temperature independent spin relaxation process was observed and attributed to quantum spin relaxation, we observed no such behavior in $Dy_2Ge_2O_7$.

Single crystal and polycrystalline powder $Dy_2Ge_2O_7$ samples were prepared using flux growth and standard solid-state synthesis techniques respectively [21], and all results reported were qualitatively consistent between the two types of samples. Powder X-ray diffraction confirmed all samples to be single phase and to have a tetragonal structure. Detailed structural analysis of the pyrogermanates ($Ho_2Ge_2O_7$, $Er_2Ge_2O_7$, $Dy_2Ge_2O_7$) has shown that the sublattice of the magnetic ions can be regarded as a spiral of alternating edge-sharing and corner-sharing triangles. Fig. 1(a) shows the rare earth lattice structure with the crystallographic axes marked (c-axis is along the [001] direction) [21]. DC magnetization measurements were performed using a Quantum Design Superconducting Quantum Interference Device (SQUID) magnetometer. We also performed AC magnetic susceptibility measurements with the ACMS option of the Quantum Design PPMS cryostat. Specific heat measurements using a standard semiadiabatic heat pulse technique were carried out in the PPMS cryostat with the $He^3$ option. The powder samples for specific heat measurements were pressed with Ag to facilitate thermal equilibration and the specific heats of Ag and the addenda were measured separately and subtracted.

Fig. 1(b) shows the temperature dependence of the DC susceptibility (*M/H*) for a powder sample in a range of different magnetic fields. The susceptibility has a peak at ~ 2.3 K in small applied fields, and the peak is broadened and moved to lower temperatures with increasing field. The measured susceptibility is the same for both field-cooled and zero-field-cooled samples, which is consistent with the peak being associated with a transition from a paramagnetic to a long-range-ordered antiferromagnetic state. This is consistent with a previous report that $Dy_2Ge_2O_7$ develops magnetic ordering at around 2.15 K [22], although detailed information about the nature of the magnetic interactions, magnetization, and other specifics are not available. With increasing applied magnetic field, the Zeeman energy associated with the external field increases, competing with the antiferromagnetic exchange energy resulting in the observed broadening and suppression of the peak in susceptibility. The inset to Fig. 1(b) shows the expanded view of the Curie-Weiss fit of the inverse susceptibility in a 0.01 T field from 100 K to 250 K, giving a Weiss temperature $\theta_w$ of -4.4 K and an effective moment of 10.5 $\mu_B$ per Dy ion, close to both the previously reported [11] and theoretical values.

The appearance of long range antiferromagnetic order at low temperature is also confirmed by the magnetic specific heat measurements on powder samples, as shown in Fig. 2(a). The phonon contributions to the specific heat were subtracted out by fitting the data between $T$ = 10 - 20 K to the Debye $T^3$ - law. In zero applied field, a sharp λ-shaped peak is observed in the temperature dependence of the magnetic heat capacity, indicating the formation of long range magnetic order. The sharp peak gradually broadens and vanishes with increasing applied magnetic field, suggesting that long range order is replaced by shorter range antiferromagnetic correlations. As was observed in the DC susceptibility measurements shown in Fig. 1(b), the peak position indicative of the phase transition shifts to lower temperatures with increasing magnetic field, characteristic of an antiferromagnetic ground state. The integrated magnetic entropy from the lowest temperature measured to 18 K is shown in the inset to Fig. 2(a). The integrated zero field entropy is about 5.2 J/mol·K by 18 K, about 90% of Rln2 and considerably larger than that of the canonical spin ice $Dy_2Ti_2O_7$ in which the well known 'zero point entropy' suppresses the measured entropy [6]. The entropy at $T_N$ is much smaller than Rln2 (about 53%), suggesting the development of short-ranged correlations above the phase transition. As in the spin ice materials, application of a magnetic field increases the entropy to Rln2, a strong indication that the $Dy^{3+}$ ions in $Dy_2Ge_2O_7$ behave as Ising spins (presumably due to crystal electric fields, as in the spin ice case). The slightly reduced measured entropy in low fields may be associated with quenched disorder in the boundaries between domains of different orientations of the antiferromagnetic ordering or the magnetic frustration resulting from the triangular sublattice of $Dy^{3+}$ spins.

Measurements of the isothermal magnetization as a function of applied field, $M(H)$ at $T = 1.8$ K are shown in Fig. 2(b) for a powder sample. For comparison, we also plot the corresponding isothermal curve of the spin ice material $Dy_2Ti_2O_7$. Interestingly, the isothermal curves of the two materials almost overlap each other, despite the differences in the crystal structures. The saturation magnetization is about 5.0 $\mu_B$ per Dy ion, about half the expected full moment for a free $Dy^{3+}$ ion. These data again suggest that the $Dy^{3+}$ ions in $Dy_2Ge_2O_7$ possess an Ising-like doublet ground state since one obtains the half moment from angular averaging in a powder sample [23]. This is in good agreement with both the integrated entropy data discussed above and shown in Fig. 2(a), and previous theoretical predictions that the low temperature ground state of the $Dy_2Ge_2O_7$ is a Kramers doublet separated by about 161 K from the first excited state due to the strong crystal electric field [12,16].

We also measured the field dependent magnetization on a $Dy_2Ge_2O_7$ single crystal sample with magnetic field applied both in plane (Fig. 3a) and out-of-plane (i.e, [001]) (Fig. 3b). We find that the magnetization at $H = 7$ T along the [001] direction is much smaller than that along the in-plane directions and far from being saturated, suggesting strong magnetic anisotropy with the magnetic hard axis pointing along the [001] direction, consistent with previous reports [12]. Similar results reported for isostructural $Ho_2Ge_2O_7$ [21] are believed to be due to the strong crystal electric field anisotropy. In Fig. 3(a), we also see that the saturation magnetization along the [110] direction is larger than that along the [100] direction, indicating that [110] is the easy axis. This is consistent with the spiral magnetic spin configuration determined from neutron diffraction experiments on $Ho_2Ge_2O_7$ [21]. Careful examination of the $M(H)$ data at low field reveals a distinct change in the slope of $M(H)$ along both in-plane directions and also in the powder sample (see inset to Fig. 2(b)) near $H = 0.06$ T while around 0.7 T along the c-axis. This feature presumably corresponds to a field-induced spin-flop transition out of the long-range-ordered antiferromagnetic state.

Although there are differences in the crystal structure, the similarity in the chemical formula and the single ion ground state of $Dy_2Ge_2O_7$ and $Dy_2Ti_2O_7$ leads one to wonder if the exotic spin relaxation processes seen in the latter system [19] are also manifested in $Dy_2Ge_2O_7$. Fig. 4(a) shows the temperature dependence of the real part of AC susceptibility $\chi'$ of $Dy_2Ge_2O_7$ powder with different applied fields at a characteristic frequency of $f = 1$ kHz. There are two features of interest. First, in zero applied field, there is a strong sharp peak in $\chi'(T)$ at around $T = 2.3$ K, which corresponds to the magnetic transition seen in the DC susceptibility measurement (open symbols). Increasing the value of the applied field significantly reduces this peak due to the suppression of the antiferromagnetic long range order. Second, a higher temperature peak appears in $\chi'(T)$ with the application of a magnetic field larger than 0.4 T. This is reminiscent of the canonical spin ices, $Dy_2Ti_2O_7$ [18,19], $Ho_2Ti_2O_7$ [20] and $Dy_2Sn_2O_7$ [8], where a peak in $\chi'(T)$ at $T \sim 15$ K has been observed and attributed to the freezing of thermally-induced single ion spin fluctuations. In contrast to $Dy_2Ti_2O_7$ and $Dy_2Sn_2O_7$ where no external field is required to observe this freezing, the higher temperature peak in $Dy_2Ge_2O_7$ is only present in an applied magnetic field, similar to the case of $Ho_2Ti_2O_7$ [20]. Fig. 4(b) shows the field dependence of both the low temperature (peak 1) and high temperature (peak 2) peak positions, with the onset of the high temperature peak only observed for fields larger than 0.4 T Both peak positions monotonically increase with increasing field and are suppressed at the highest fields when the magnetization saturates.

Focusing on the low field AC susceptibility, Fig. 5 shows $\chi'(T)$ of $Dy_2Ge_2O_7$ powder samples measured at several characteristic frequencies at $H = 0$, 0.11, and 0.20 T. We observe a frequency dependence for $H = 0$ only for our highest measurement frequency (10 kHz). However, when the external field suppresses the long range order (above 0.06 T), the peak shifts to higher temperature and broadens with increasing frequency. As discussed above, this material is not a spin glass at low temperatures – as evidenced by the sharp peak in specific heat and the lack of bifurcation between the field-cooled and zero-field-cooled DC susceptibility. We speculate that our observed frequency dependence is associated with the collective relaxation of short-range correlated spins and the possible presence of antiferromagnetic domains in zero field. Neutron scattering experiments on isotopically-enriched $Dy_2Ge_2O_7$ could address this issue..

To investigate the higher field AC susceptibility, in Fig. 6(a) we plot the $\chi'(T)$ at different frequencies in a magnetic field of 0.5 T. Taking both peak temperature positions from the $\chi'$ curves for different frequencies we performed Arrhenius fits of the form $f = f_0 e^{-E_A/k_B T}$ as shown in the inset, where $E_A$ is the energy barrier to spin relaxation and $f_0$ is the effective attempt frequency. From the fit to the high temperature peaks we find $E_A$ to be about 162 K, very close to the calculated values for the first excited state due to the crystal field splitting [16]; $f_0$ is roughly 30 MHz. A fit to the low temperature peaks gives $E_A$ about 33 K and $f_0$ about 6 MHz. These values are physically very accessible for this system, strongly suggesting that both low and high temperature peaks are associated with thermal spin relaxation processes.

Fig. 6(b) shows the frequency dependence of the imaginary part of AC susceptibility, $\chi''(f)$, measured at different temperatures in a 0.5 T external field. This sort of Casimir -du Pré plot shows a peak at characteristic spin relaxation frequencies [19,24]. Interestingly, the data for temperatures below 8 K are suggestive of a double-peak structure to $\chi''(f)$, indicating that two different spin relaxation mechanisms are relevant. This feature has previously been observed in $Dy_2Ti_2O_7$ in a similar magnetic field. In that case, however, one peak is independent of temperature and can be associated with a quantum spin relaxation while the other peak shifts to higher frequency with increasing temperature and is associated with thermally induced spin relaxation [19]. In $Dy_2Ge_2O_7$, both peaks shift to higher frequencies with increasing temperature, indicating the absence of quantum relaxation process. This suggests the presence of two thermal relaxation processes, perhaps one associated with single spin relaxation and the other associated with the collective relaxations of correlated clusters of spins, consistent with our interpretation of the presence of the two maxima in $\chi'(T)$. An analogous pair of thermal relaxation phenomena is observed in $Dy_2Ti_2O_7$, associated with single-spin relaxation at higher temperatures and spins with ice-like correlations at the lowest temperatures, but with a quantum spin relaxation regime observed at intermediate temperatures [19]. The absence of quantum spin relaxation in $Dy_2Ge_2O_7$, which allows us to clearly observe the two different thermal processes, is presumably due to the difference in the crystal field environment associated with the different lattice structures.

In summary, our investigation of $Dy_2Ge_2O_7$ demonstrates that this Ising system which displays long range order in low magnetic fields shows spin freezing, associated both with single spin relaxation and with the relaxation of correlated groups of spins. Such behavior had previously been observed only in the spin ice materials, although in those materials a quantum single-spin relaxation mechanism was also evident. These results suggest that such spin relaxation phenomena are more general in rare earth magnets.

We acknowledge the financial support from NSF grant DMR-0701582. The work at Princeton was supported by NSF grant DMR-02-13706.

FIGURE CAPTIONS

Figure 1: (a) Schematic diagram of the rare earth lattice structure of $Ln_2Ge_2O_7$. The shaded areas represent the edge-sharing triangle of rare-earth ions and atoms of different color represent atoms in four crystallographic *ab* plane (Ref. 21). (b) Temperature dependence of the DC susceptibility of a $Dy_2Ge_2O_7$ powder sample with different applied magnetic fields. Inset shows the expanded view of the inverse DC susceptibility as a function of temperature and the Curie-Weiss fit (red line).

Figure 2: (a) Magnetic heat capacity $C_{mag}/T$ as a function of temperature at different fields. Inset shows the integrated entropy. (b) Field dependence of magnetization of $Dy_2Ge_2O_7$ and $Dy_2Ti_2O_7$ powders at $T = 1.8$ K. Inset is an expanded view of the region of low field.

Figure 3. Field dependence of the magnetization of a $Dy_2Ge_2O_7$ single crystal at $T = 1.8$ K with the field applied along (a) [100] (black curve) and [110] (red curve) and (b) [001] respectively. Inset is an expanded view of the region of low field.

Figure 4: (a) Temperature dependence of the real part of the AC susceptibility $\chi'$ with different applied fields at a characteristic frequency $f = 1$ kHz. Inset is an expanded view. (b) Temperature values of the low temperature peak (1) and high temperature peak (2) shown in Fig. 3(a) as a function of magnetic field.

Figure 5. Temperature dependence of the real part of the AC susceptibility χ′ at different frequencies, with 0, 0.11, and 0.20 T applied fields.

Figure 6: (a) Temperature dependence of the real part of the AC susceptibility χ′ at the frequency of 10 Hz (black), 100 Hz (red), 1 kHz (green), and 10 kHz (blue). The inset shows the Arrhenius fits as described in the text. (b) Frequency dependence of the imaginary part of the AC susceptibility χ″ at different temperatures. The measurements were performed with 0.5 T applied field.

Figure 1.
X. Ke, *et al.*

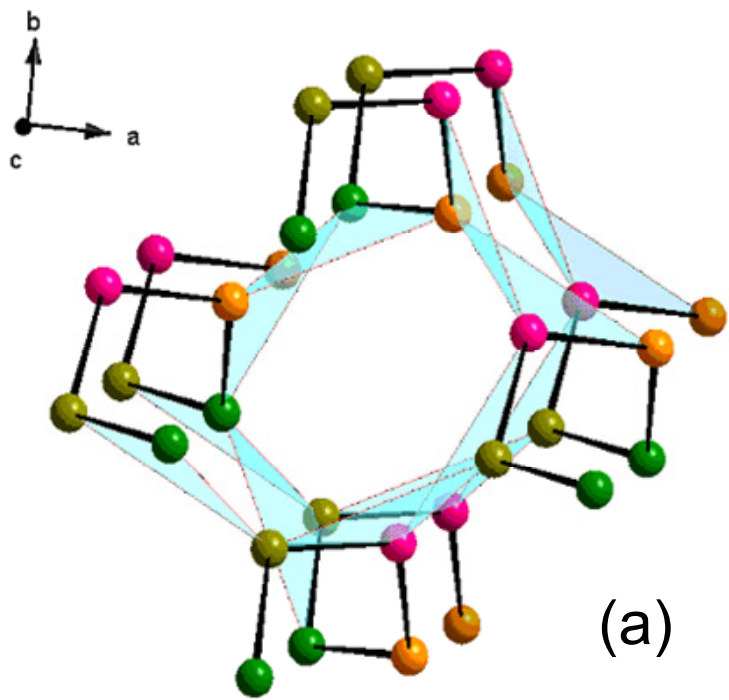

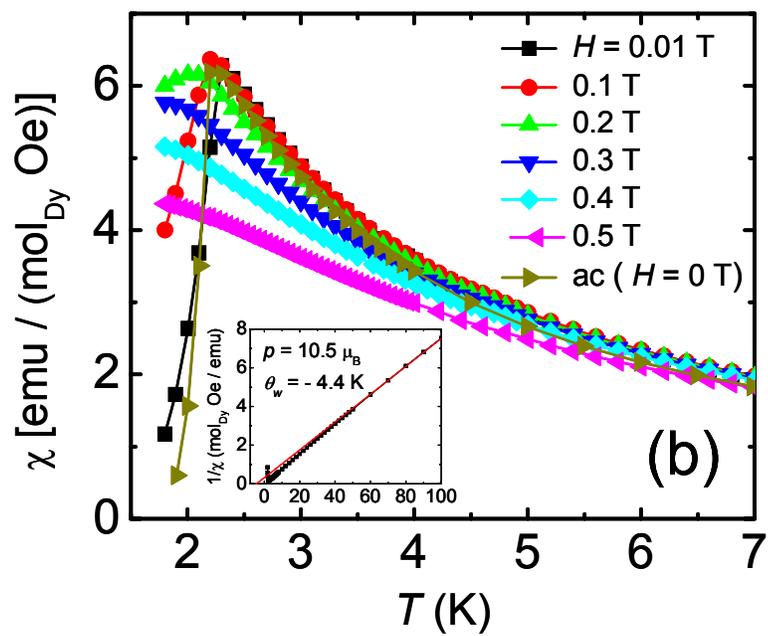

Figure 2.
X. Ke, *et al.*

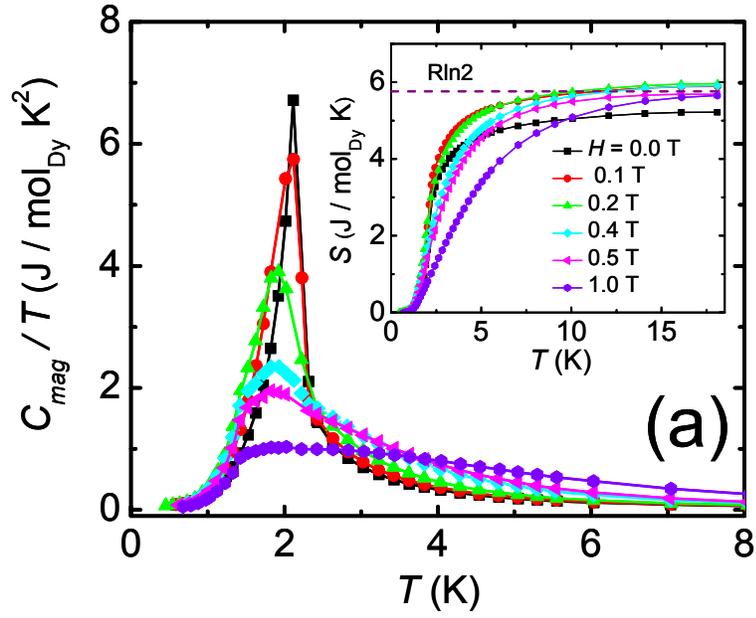

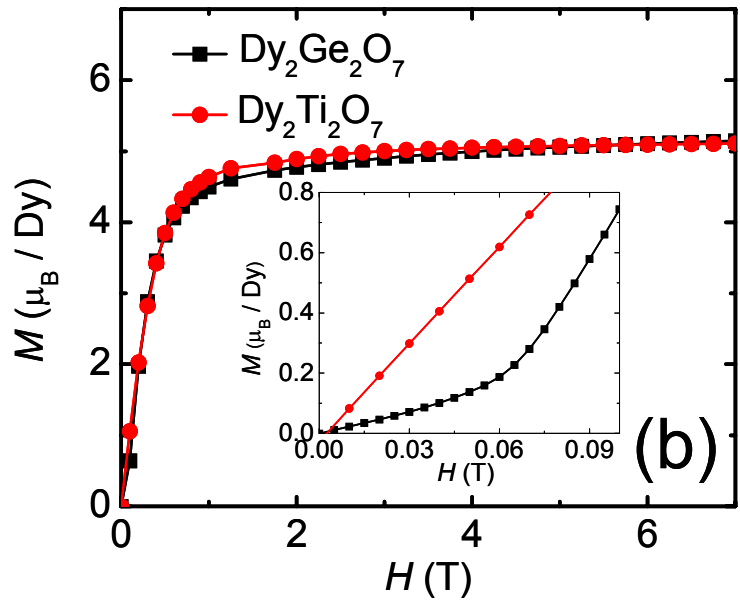

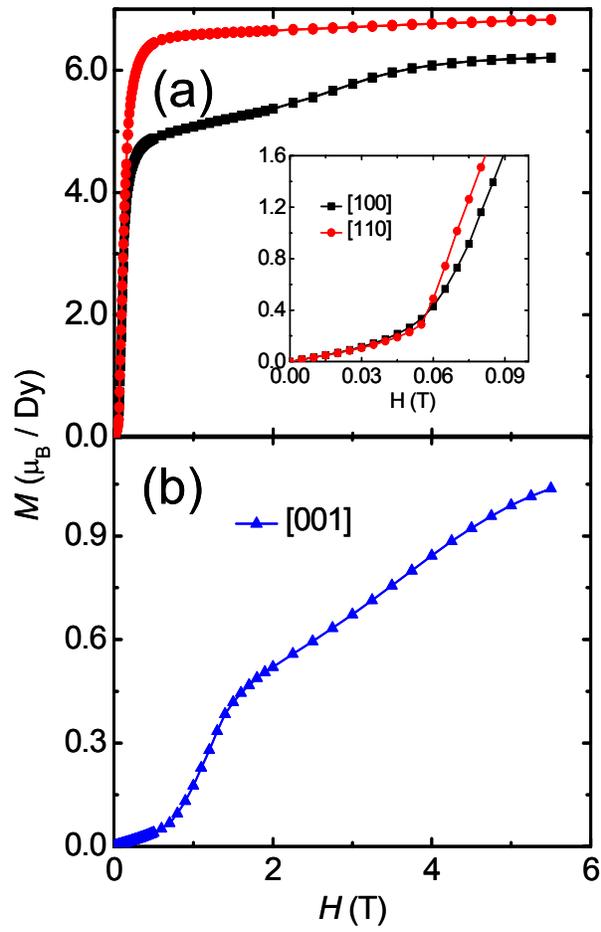

Figure 4.
X. Ke, *et al.*

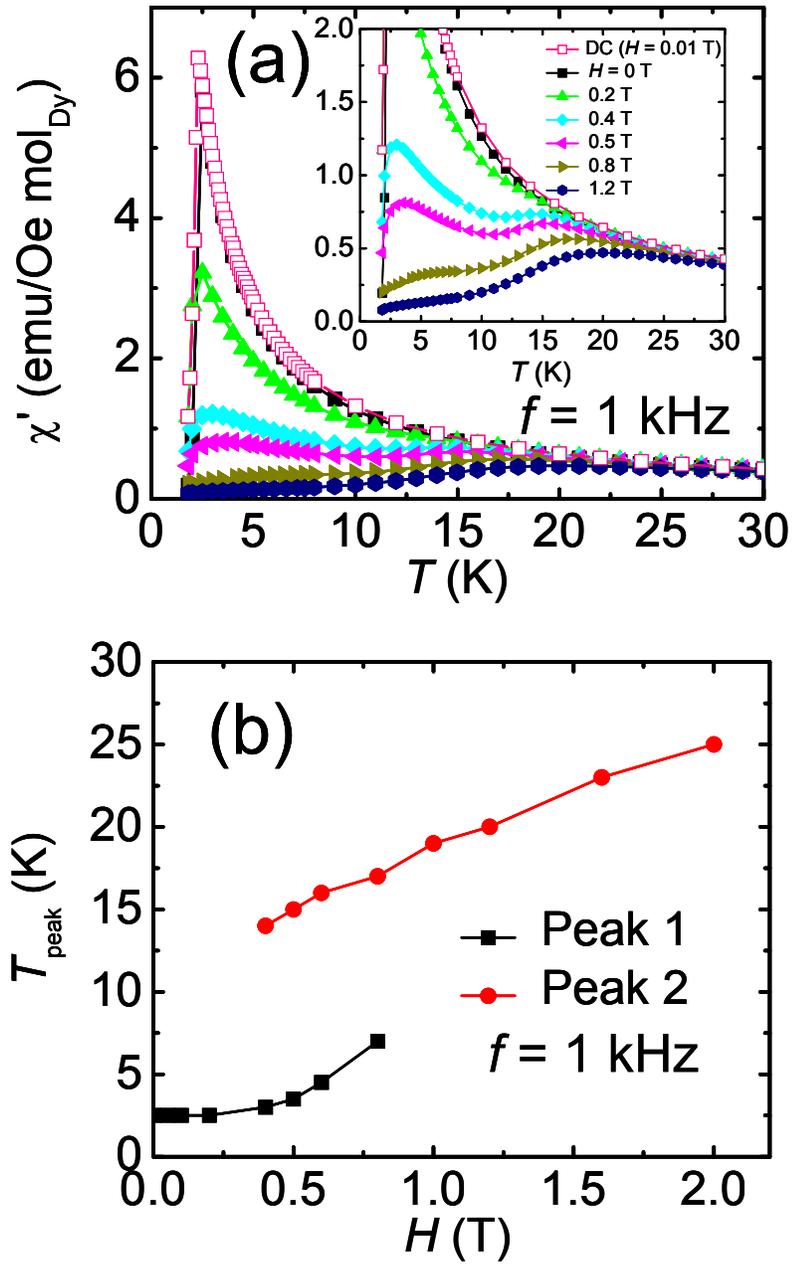

Figure 5.
X. Ke, *et al*.

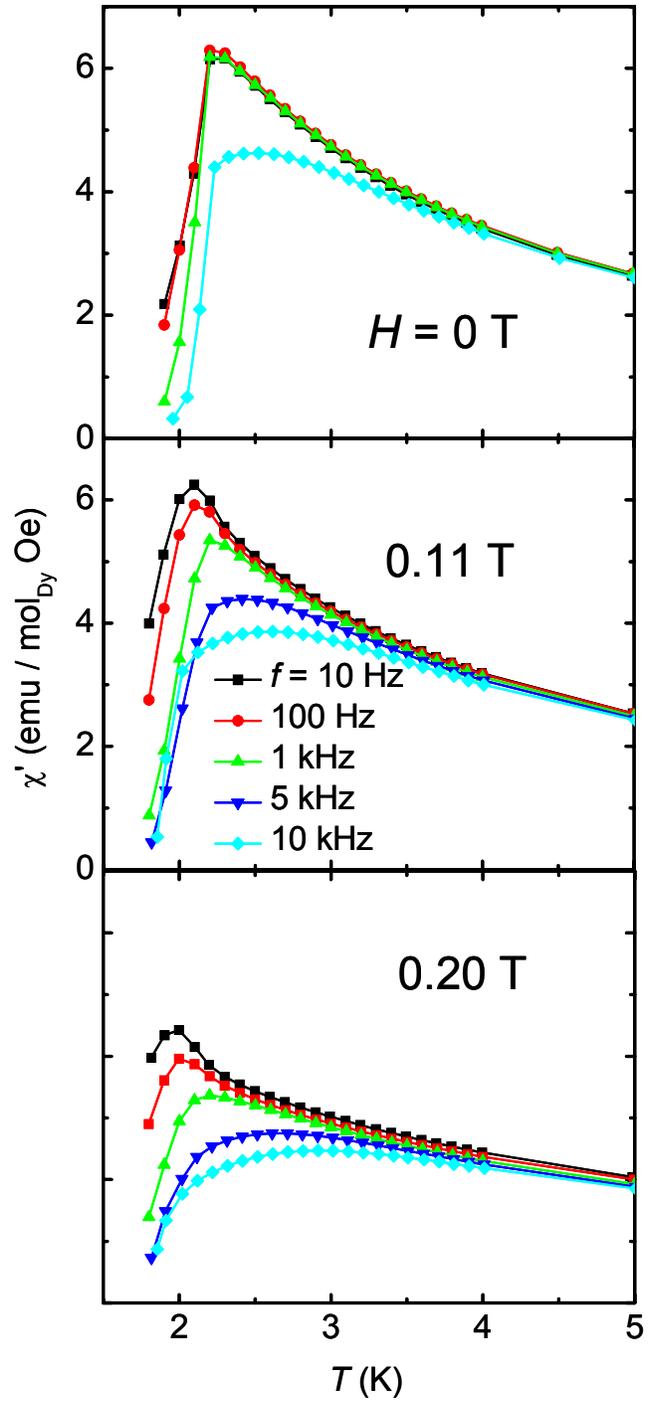

Figure 6.
X. Ke, *et al*.

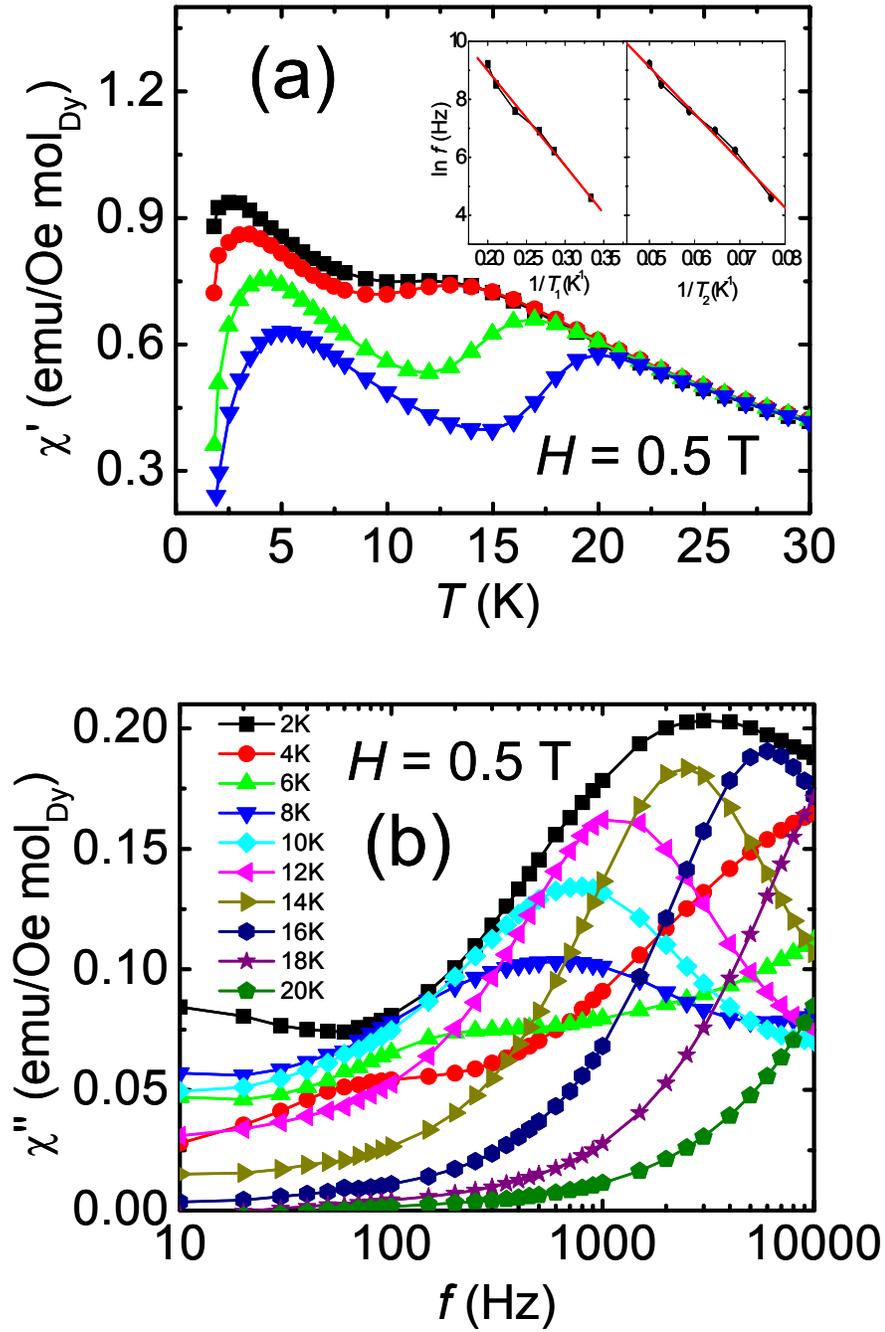